# Towards Realistic Artificial Benchmark for Community Detection Algorithms Evaluation

Günce Keziban Orman*[1,2]

Vincent Labatut[1]

Hocine Cherifi[2]

[1]Galatasaray University, Computer Science Department
Çırağan Cad. No:36, Ortaköy 34357, Istanbul, Turkey

[2]LE2I  UMR CNRS 6306
Université de Bourgogne, Faculté des Sciences Mirande
9, avenue Alain Savary BP 47870, 21078, Dijon, France

*Corresponding author

**Abstract:** Many algorithms have been proposed for revealing the community structure in complex networks. Tests under a wide range of realistic conditions must be performed in order to select the most appropriate for a particular application. Artificially generated networks are often used for this purpose. The most realistic generative method to date has been proposed by Lancichinetti, Fortunato and Radicchi (LFR). However it does not produce networks with some typical features of real world networks. To overcome this drawback, we investigate two alternative modifications of this algorithm. Experimental results show that in both cases, centralization and degree correlation values of generated networks are closer to those encountered in real-world networks. The three benchmarks have been used on a wide set of prominent community detection algorithms in order to reveal the limits and the robustness of the algorithms. Results show that the detection of meaningful communities gets harder with more realistic networks, and particularly when the proportion of inter community links increases.

**Keywords:** Community structure, Topological Properties, LFR benchmark, Configuration Model, Preferential Attachment.

## 1  Introduction

Because of the spread of complex network applications, the community detection problem has been studied in many different areas such as computer science, biology, sociology, resulting in numerous algorithm based on a whole range of principles (Fortunato 2010). In order to perform community detection on a specific real-world network, one needs to select the most appropriate tool. This choice is difficult because of the profusion of methods, and also of the variability of their performance according to the networks characteristics. Being able to compare them therefore becomes a very important methodological need.

Most of these algorithms represent the community structure under the form of a node partition. Their performance can consequently be assessed by comparing an estimated partition to a reference partition, the latter representing the known community structure of the considered network. The question remains to determine which data should be used as a benchmark. Real-world networks are very heterogeneous and not so numerous. It is thus difficult to select a collection of such networks matching the topological properties of the targeted system.

Artificial networks seem to be an appropriate alternative. They are widely used to compare community detection algorithms performances (Danon et al. 2005; Orman et al. 2011a). Indeed, generative models allow producing easily and quickly large collections of such networks. Moreover, these models provide a certain control on the topological properties of the generated networks, making it possible to mimic the targeted system features. The only point of concern is the level of realism of the generated networks, which is a prerequisite to obtain relevant test results. For this purpose, generative models are generally defined in order to reproduce known real-world networks properties. Of course, current knowledge regarding these properties may not be exhaustive, and we can consequently never be completely assured the generated networks are perfectly realistic. For this reason, tests on artificial networks should be seen as complementary to tests on real-world networks.

Up to now, a few methods have been designed to generate networks with a community structure. The most popular one is certainly the model by Girvan and Newman (GN) (Girvan and Newman 2002). Although widely used to test and compare algorithms (Donetti and Munoz 2004; Duch and Arenas 2005; Girvan and Newman 2002; Radicchi et al. 2004), it is limited in terms of realism (Lancichinetti et al. 2008): the generated networks are rather small compared to most real-world networks; all nodes have roughly the same degree; and all communities have the same size. Yet, typically both the community size distribution and the degree distribution follow a power law in real-world complex networks (da Fontura Costa et al. 2008; Guimerà et al. 2003). To tackle this problem, several GN variants have been defined, producing larger networks, and communities with heterogeneous sizes (Danon et al. 2006; Fortunato 2010; Pons and Latapy 2005).

More recently, a different approach appeared, based on rewiring: first an initial network with desired properties (but no community structure) is randomly generated, then virtual communities are drawn, and finally some links are rewired so that these communities appear in the network. The method described by Bagrow (Bagrow 2008) uses the Barabási–Albert model (Barabasi and Albert 1999) to generate the initial network, resulting in a power law degree distribution, but produces small networks with equal-sized communities. The method by Lancichinetti *et al.* (LFR) (Lancichinetti et al. 2008) is based on the configuration model (Molloy and Reed 1995), which generates networks with power law degree distribution, too. However, unlike Bagrow's method, LFR generates power law distributed community sizes, and the network size is not constrained. Although LFR exhibits the most realistic properties, it also has some noticeable limitations. Previous experiments showed the generated networks exhibit a low transitivity and close to zero degree correlation for certain community structures (Orman and Labatut 2009), while according to Newman (Newman 2003), real-world networks usually have a clearly non-zero degree correlation, and their transitivity is relatively high.

Interestingly, improvements on the realistic aspect of the generated networks have a noticeable effect on most community detection algorithms behaviour. A performance drop was observed when authors switched from equal-sized communities to heterogeneous distributions (Danon et al. 2006; Pons and Latapy 2005). The introduction of a power law degree distribution also made the benchmarks more discriminatory, allowing to highlight differences between algorithms whose performances were considered similar before (Lancichinetti et al. 2008).

In this work, we study the impact of network realism on the partitioning performance of community identification algorithms. We propose and evaluate two modifications of the LFR method to improve the realism of the generated networks. The realism level is appreciated by comparing the main known topological properties of the generated networks with some reference values commonly observed in real-world networks. In order to assess the influence of variations in the realism level, eleven representative community detection algorithms are tested on artificial networks generated by the original and modified LFR methods.

The rest of this article is organized as follows. Section 2 describes the topological properties generally used to characterize complex networks. Section 3 is dedicated to present the methods used to generate the networks. We first briefly define the LFR method and its characteristics, then our proposed modifications. Section 4 is a short description of the community detection algorithms used to test the effect of network realism on partitioning performance. We present the benchmark data, the results regarding the realism of the generated networks and its effect on community detection in section 5. Finally, in section 6, we highlight our contributions and propose some further extensions of our work.

## 2 Topological Properties

Undirected real-world networks are known to share some common properties. In this section, we present the most prominent ones: small-worldness, transitivity, degree-related properties and centrality-related properties. Many other properties can be used to describe a network, either by analysing some measure, like network diameter (Boccaletti et al. 2006), or by counting the number of occurrences of a given substructure like motifs in (Milo et al. 2002). But their use is not really widespread, and we would consequently lack experimental values to take advantage of them in this work.

**Small-Worldness.** A model is said to have the small-world property if, for a fixed average degree, the average distance (i.e. the length of the shortest path) between pairs of nodes increases logarithmically with the number of nodes $n$ (Newman 2003). This property is important, because it is related to the network efficiency to propagate information.

**Transitivity.** The transitivity property is measured by a transitivity coefficient $C$, also called clustering coefficient (Watts and Strogatz 1998). Two different versions exist, both trying to assess the density of triangles in a network, but in slightly different ways. The higher this coefficient, the more probable it is to observe a link between two nodes having a common neighbour. Independently of the considered coefficient version, a real-world network is supposed to have a higher transitivity than a Poisson random network (such as those generated by the Erdős–Rényi model (Erdos and Renyi 1959)) with the same number of nodes and links, by a factor of order $n$ (Newman 2003).

**Degree Distribution.** Networks can also be described according to their degree distribution. In most real-world networks, this distribution follows either a power or an exponential law. In other words, the probability for a node to have a degree $k$ is either $p_k \sim k^{-\gamma}$ or $p_k \sim e^{-k/K}$ (Newman 2003). Networks with a power-law degree distribution are the most common. They are called scale-free, because their degree distribution does not depend on their size (some other properties may, though). Experimental studies showed that the $\gamma$ coefficient usually ranges from 2 to 3 (Barabasi and Albert 2002;

Boccaletti et al. 2006; Newman 2003).

**Average and Maximum Degree.** In a real-world network, the average and maximal degrees generally depend on the number of nodes it contains. For a scale-free network, it is estimated to be $\langle k \rangle \sim k_{max}^{-\gamma+2}$ (Barabasi and Albert 2002; Boccaletti et al. 2006) and $k_{max} \sim n^{1/(\gamma-1)}$ (Newman 2003), respectively.

**Degree Correlation.** The degree correlation of a network constitutes another interesting property. It indicates how a node degree is related to its neighbours'. Real-world networks usually show a non-zero degree correlation. If it is positive, the network is said to have assortatively mixed degrees, whereas if it is negative, it is disassortatively mixed (Newman 2003). According to Newman, social networks tend to be assortatively mixed, while other kinds of networks are generally disassortatively mixed. Nodes with high degree are called hubs, because they have a more central role in the network.

**Centralisation.** A centrality measure determines how influential a node is within a network. Among the various existing definitions, degree, closeness and betweenness centralities as described in (Freeman 1979) are the most widely used. *Degree centrality* measures the involvement of a node in the network by the number of nodes connected to it. This local definition does not take into account the position of the node in the network and therefore cannot measure the ability of the node to reach others quickly. *Closeness centrality,* defined as the inverse sum of shortest distances to all the other nodes of the network, captures this feature. *Betweenness centrality* asserts the ability of a node to plays a broker role in the network, by measuring the degree to which it lies on the shortest path between other nodes.

While centrality measures the leadership of a node, centralisation is global: it expresses how much the network is organised around some structurally important nodes. A very centralised network is dominated by one or a few very central nodes, and takes the form of a star or a wheel. It is very sensitive to failures or attack on those nodes, whereas a less centralised network (e.g. completely connected), is more resilient. Centralisation measures are based on the differences between the centrality of the most central node and those of all the other ones. Its definition is general, so it can be based on any of the three previously presented centrality measures.

## 3 Network Generation

The LFR method was proposed by Lancichinetti *et al.* (Lancichinetti et al. 2008) to randomly generate networks with mutually exclusive communities. The generation process is two-stepped. First, a network is produced by using a scale-free random network model. Second, a community structure is randomly drawn, and the network is rewired to make it appear. In this study, we do not change the rewiring procedure but we substitute another random model to the original one used in the first step. In this section, we first describe the genuine LFR method, then the two alternative models we considered.

*3.1 Original LFR Method*

The original LFR method uses the configuration model (Molloy and Reed 1995) in its random network generation step, in order to generate a network containing $n$ nodes, with average degree $\langle k \rangle$, maximum degree $k_{max}$ and a power law distributed degree with exponent $\gamma$. Then, the rewiring process is applied in two phases. First, the communities are randomly drawn, so that their sizes follow a power law distribution with exponent $\beta$. These are just virtual communities, i.e. they are just groups of nodes, and the topology of the network does not reflect them for now. Second, an iterative process takes place to rewire certain links, so that the community structure appears. It consists in moving a predefined proportion of the network links inside the communities, but without changing the node degree. This proportion is specified using a parameter called the *mixing coefficient* $\mu$.

The mixing coefficient represents the desired average proportion of links between a node and nodes located outside its community. It is generally not possible to meet this constraint exactly, and the mixing coefficient is therefore only approximated in practice. Its value determines how clearly the communities are defined. For small $\mu$ values, the communities are distinctly separated because they share only a few links, whereas when $\mu$ increases the proportion of inter community links becomes higher, making community identification a difficult task. The network has no community structure for a limit value of the mixing coefficient given by: $\mu_{lim} > (n - n_c^{max})/n$, where $n$ and $n_c^{max}$ are the number of nodes in the network and in the biggest community, respectively (Lancichinetti and Fortunato 2009b).

The LFR method was subsequently extended to generate weighted and/or oriented networks, with possibly overlapping communities (Lancichinetti and Fortunato 2009a, b). However, our focus is on undirected and unweighted networks, so we based our work on the original version (Lancichinetti et al. 2008). It guaranties obtaining several realistic properties: size of the network, power law distributed degrees and community sizes. Moreover, some parameters give the user a direct control on these properties: network size ($n$), degree distribution ($\gamma$, $k_{max}$, $\langle k \rangle$), community structure ($\beta$, $\mu$). However, there is no direct control on the other properties, such as those presented in section 2: small-worldness, transitivity, degree correlation, node centrality and centralization.

*3.2 Modified LFR Method*

The configuration model is very flexible as it is able to produce networks with any size and degree distribution, but it is known to generate networks with zero correlation (Serrano and Boguñá 2005) and low transitivity when degrees are power law distributed (Newman 2003). This is why we propose to replace it by some generative model, with more realistic properties. We considered the Barabási–Albert *preferential attachment* model (Barabasi and Albert 2002) and one of its variants called *evolutionary preferential attachment* (Poncela et al. 2008). Both methods allow generating scale-free networks with desirable size and average degree. Furthermore as we still use the LFR rewiring process, community sizes stay power law distributed with exponent $\beta$.

The Barabási–Albert preferential attachment model (BA) (Barabasi and Albert 1999) was designed as an attempt to explain the power law degree distribution observed in real-world networks by the building process of these networks. Starting from an initial network containing $m_0$ connected nodes, a realistic iterative process is applied to simulate growth. At each iteration, one node is added to the network, and is randomly connected to $m$ existing nodes ($m \leq m_0$). These $m$ nodes are selected with a probability which is a function of their current degree $k$: $P(k_i) = k_i / \sum_j k_j$. In other words: the higher a node degree, the higher its chances of being selected. This so-called preferential attachment mechanism results in a power law degree distribution, since degree increases faster for nodes with higher degree, as new nodes are added to the network. The exponent $\gamma$ of the power law cannot be controlled though, and tends towards 3 (Barabasi and Albert 1999). The average degree depends directly on the $m$ parameter: $\langle k \rangle = 2m$ (Newman 2003). The average distance is always less than in same-sized Erdős-Rényi networks, so it has the small world property (Barabasi and Albert 2002).Transitivity is greater than in Erdős-Rényi networks, but nevertheless decreases with network size following a power law $C \sim n^{-0.75}$ (Barabasi and Albert 2002).

The evolutionary preferential attachment (EV) (Poncela et al. 2008) model is a variant of the BA model. It also uses the preferential attachment and growth mechanisms, except the attachment probabilities are not based on some topological properties, like the current degree in the case of BA, but on some nodal dynamic property, updated using the prisoner's dilemma game. Every few iterations, each node plays either cooperation or defection against all its neighbours. It gets a total score depending on the individual results: 0 for unilateral cooperation or bilateral defection, 1 for bilateral cooperation, and $b$ for unilateral defection, with $b > 1$. The first move is randomly chosen, whereas the next one depends on the respective results of the considered node and a randomly picked neighbour. If the neighbour's score is better, the node might switch to its strategy, with a probability depending on the difference between their scores. Nodes with higher scores are more attractive to a node added to the network, because by being connected to them, it may use a strategy which proved to be successful. According to its authors, this process is more realistic and leads to networks with high transitivity and degree correlation. Besides the parameters already needed by BA ($n$, $m_0$ and $m$), EV uses $b$ (points scored for unilateral cooperation) and ε (selection pressure). The latter allows to modulate the influence of the preferential attachment mechanism: all nodes are equiprobable when $\varepsilon = 0$, whereas the nodes scores are fully considered for $\varepsilon = 1$.

As the generating processes differ only in the first step of the LFR algorithm, for simplicity matters we will thereafter refer to the network generators by using the name of the model employed during the first step. Consequently, LFR-CM will correspond to the original LFR method, whereas LFR-BA and LFR-EV are the modified versions based on the corresponding models.

## 4 Community Detection Algorithms

Over the years, many methods have been devised to provide efficient community detection algorithms. As the spectrum is wide, building of taxonomy of solutions is not trivial. Each approach can be affected differently by the level of realism of the networks, so it is necessary to select a representative set of algorithms to apply on our benchmark. In this section, we present the different categories we identified, and the representative set of algorithms we selected for evaluation.

*4.1 Link-Centrality-Based Algorithms*

The algorithms based on link-centrality measures rely on a hierarchical divisive approach. Initially the whole network is seen as a single community, i.e. all nodes are in the same community. The most central links are then repeatedly removed. The underlying assumption is that these particular links are located between the communities. After a few steps, the network is split in several components which can be considered as communities in the initial network. Iterating the process, one can split each discovered community again, resulting in a finer community structure. Algorithms of this category differ in the way they select the links to be removed. The first and most known algorithm using this approach was proposed by Newman (Girvan and Newman 2002), and relies on the *edge-betweenness* measure. It estimates the centrality of a link by considering the proportion of shortest paths going through it in the whole network. As the complexity of this algorithm is high, it is not well suited for very large networks. Radicchi *et al.* proposed a variation called *Radetal* (RA) (Radicchi et al. 2004), based on *link transitivity* instead of edge-betweenness. This measure is defined as the number of triangles to which a given link belongs, divided by the number of triangles that might potentially include it. Its lower complexity makes it more appropriate for large networks. It is used as the representative of the link centrality based approach.

*4.2 Modularity Optimization Algorithms*

Modularity is a prominent measure of the quality of a community structure introduced by Newman and Girvan (Newman and Girvan 2004). It measures the internal connectivity of the identified communities, relatively to a randomized null model. Modularity optimization algorithms try to find the best community structure in terms of modularity. They diverge on the optimization process they are based upon. As this approach is very influential in the community detection literature, we consider three algorithms for investigation.

*FastGreedy* (FG), developed by Newman *et al.* (Newman 2004), relies on a greedy optimization method to implement a hierarchical agglomerative approach. The agglomerative approach is symmetrical to the divisive one described in the previous subsection. In the initial state, each node constitutes its own community. The algorithm merges those communities step by step until only one remains, containing all nodes. The greedy principle is applied at each step, by considering the largest increase (or smallest decrease) in modularity as the merging criterion. Because of its hierarchical nature, FG produces a hierarchy of community structures like the divisive approaches. The best one is selected by comparing their modularity values.

*Louvain* (LV) is another optimization algorithm proposed by Blondel et *al.* (Blondel et al. 2008). It is an improvement of FG, introducing a two-phase hierarchical agglomerative approach. During the first phase, the algorithm applies a greedy optimization to identify the communities. During the second phase, it builds a new network whose nodes are the communities found during the first phase. The intra-community links are represented by self-loops, whereas the inter-community links are aggregated and represented as links between the new nodes. The process is repeated on this new network, and stops when only one community remains.

*Spinglass* (SG) by Reichardt and Bornholdt (Reichardt and Bornholdt 2006) relies on an analogy between a popular statistical mechanic model called *Potts spin glass,* and the community structure. It applies the simulated annealing optimization technique on this model to optimize the modularity.

*4.3 Spectral Algorithms*

Spectral algorithms take advantage of various matrix representations of networks. Classic spectral graph partitioning techniques focus on the eigenvectors of the Laplacian matrix. They were designed to find the partition minimizing the links lying in-between node groups. The methods we selected are variants adapted to complex networks analysis.

*Leading Eigenvector* (LEV) was proposed by Newman (Newman 2006). It applies the classic graph partitioning approach, but to the modularity matrix instead of the Laplacian. Doing so, it performs an optimization of the modularity instead of the objective measures used in classic graph partitioning, such as the minimal cut.

*Commfind* (CF) was developed by Donetti and Muñoz (Donetti and Munoz 2005). It combines the analysis of the Laplacian matrix eigenvectors used in classic graph partitioning with a cluster analysis step. Instead of using the best eigenvector to iteratively perform bisections of the network, it takes advantage of the *m* best ones. Communities are obtained by a cluster analysis of the projected nodes in this *m*-dimensional space.

*4.4 Random-Walk-Based Algorithms*

Several algorithms use random walks in various ways to partition the network into communities. We retain two of them in our comparisons.

*Walktrap* (WT), by Pons and Latapy (Pons and Latapy 2005), uses a hierarchical agglomerative method like FG, but with a different merging criterion. Unlike FG, which relies on the modularity measure, WT uses a node-to-node distance measure to identify the closest communities. This distance is based on the concept of random-walk. If two nodes are in the same community, the probability to get to a third one located in the same community through a random walk should not be very different for both of them. The distance is constructed by summing these differences over all nodes, with a correction for degree.

*MarkovCluster* (MCL) simulates a diffusion process in the network to detect communities (van Dongen 2008). This method relies on the network *transfer matrix*, which describes the transition probabilities for a random walker evolving in this network. Two transformations called expansion and inflation are iteratively applied on this matrix until convergence. The final matrix can be interpreted as the adjacency matrix of a network with disconnected components, which correspond to communities in the original network.

*4.5 Information-Based Algorithms*

The main idea of those approaches is to take advantage of the community structure in order to represent the network using less information than that encoded in the full adjacency matrix. We selected two algorithms from this category.

*Infomod* (IND) was proposed by Rosvall and Bergstorm (Rosvall and Bergstrom 2007). It is based on a simplified representation of the network focusing on the community structure: a community matrix and a membership vector. The former is an adjacency matrix defined at the level of the communities (instead of the nodes), and the latter associates each node to a community. The authors use the mutual information measure to quantify the amount of information from the original network contained in the simplified representation. They obtain the best partition by considering the representation associated to the maximal mutual information.

*Infomap* (INP) is another algorithm developed by Rosvall and Bergstorm (Rosvall and Bergstrom 2008). The community structure is represented through a two-level nomenclature based on Huffman coding: one to distinguish communities in the network and the other to distinguish nodes in a community. The problem of finding the best partition is expressed as minimizing the quantity of information needed to represent some random walk in the network using this nomenclature. With a partition containing few inter-community links, the walker will probably stay longer inside communities, therefore only the second level will be needed to describe its path, leading to a compact representation. The authors optimize their criterion using simulated annealing.

*4.6 Other Algorithms*

A number of algorithms do not fit in the previously described approaches. We selected the *Label Propagation* (LP) algorithm by Raghavan *et al.* (Raghavan et al. 2007), which uses the concept of node neighbourhood and simulates the diffusion of some information in the network to identify communities. Initially, each node is labelled with a unique value. Then an iterative process takes place, where each node takes the label which is the most spread in its neighbourhood (ties are broken randomly). This process goes on until convergence, i.e. each node has the majority label of its neighbours. Communities are then obtained by considering groups of nodes with the same label. By construction, one node has more neighbours in its community than in the others.

## 5 Results and Discussion

In this section we present the characteristics of the artificial networks generated using the modified LFR method described in section 3. We conduct a comparative analysis of the topological properties induced by each variant and discuss the obtained level of realism. Finally, we evaluate the impact of realism on the partitioning performances of the community detection algorithms from section 4, by applying them to our artificial benchmarks and commenting their results.

*5.1 Benchmark Generation*

In order to determine the effect of the generative parameters on the uncontrolled properties of the networks, it is necessary to consider an appropriate range of values for each parameter. Since we want our networks to be realistic, these values must be, as much as possible, consistent with what is observed in real-world networks. For this matter, we used descriptions of real-world networks measurement from the literature (Barabasi and Albert 2002; Boccaletti et al. 2006; da Fontura Costa et al. 2008; Newman 2003). But the we could not find all the information needed to set up the models, which is why we also based our choices on previous experiments in artificial networks generation (Lancichinetti et al. 2008; Orman and Labatut 2009).

The network size $n$ has a direct effect on the processing time, not only regarding the generation of networks, but even more importantly concerning the community detection task. For this reason, we selected a size of $n = 5000$ nodes, which is at the same time reasonably large and computationally tractable.

Amongst the three models, CM is the only one making it possible to control the exponent $\gamma$ of the degree distribution: for BA and EV, it is fixed to 3 by construction, as mentioned in section 3.2. This value belongs to the realistic range of this parameter, however this limitation prevents us from testing if changes in $\gamma$ affect the uncontrolled properties. In order to investigate this matter, we performed an extensive experimentation on LFR-CM, for a wide range of the parameter values, and with $\gamma$ ranging from 2 to 3. The results, displayed in Figure 1, show $\gamma$ has a negligible effect on the uncontrolled properties. The fact this parameter is blocked for BA and EV is therefore not a problem for this study.

The average degree is directly related to the network size and, in the case of scale free networks, to the degree distribution exponent. However, this dependence is quite loose, which is why it is expressed formally only in an approximate way in the literature (cf. section 2). In other words, for some fixed values of $n$ and $\gamma$, one can find real-world networks with rather different densities (Barabasi and Albert 2002; Boccaletti et al. 2006; da Fontura Costa et al. 2008; Newman 2003). We consequently selected two consensual values for the average degree: $\langle k \rangle = 15; 30$. In CM, this constraint is enforced directly, whereas in BA and EV, we used $m = 7; 15$ to reach the same result. All three models allow controlling the average degree, but only CM lets the user specify the maximal degree. In order to get comparable networks, we tuned this parameter to make its values similar to what was observed in networks generated by BA and EV. We finally used $k_{max} = 45; 90$ for CM, each value corresponding to one of the average degree values.

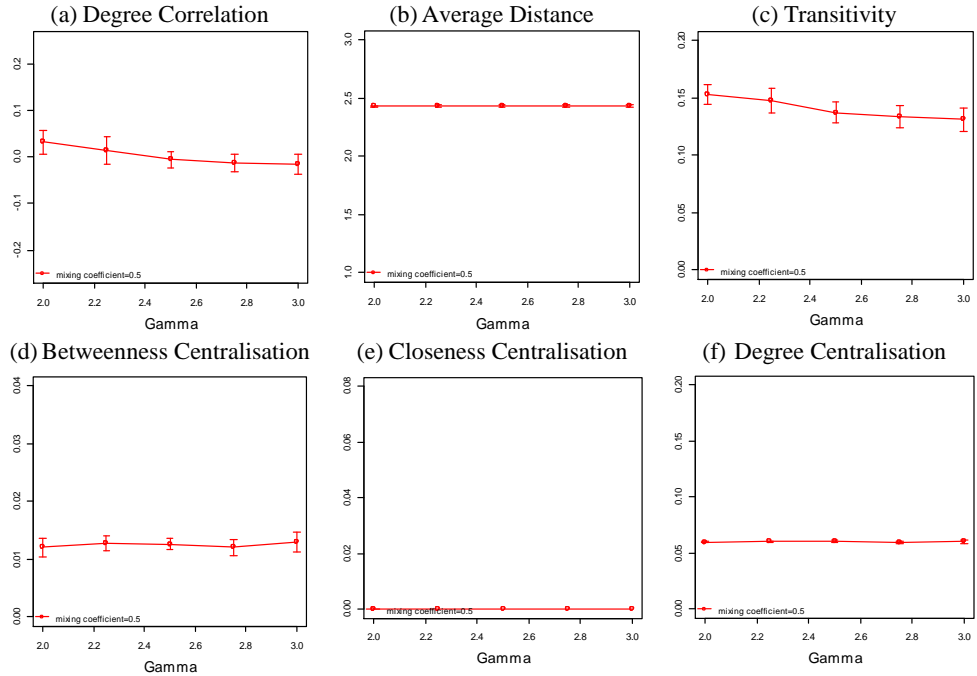

**Figure 1.** Influence of the degree distribution exponent $\gamma$ on the uncontrolled properties. Networks were generated with parameters $n = 1000, \mu = 0.5, \langle k \rangle = 30$ and using LFR-CM.

The community structure is specified by the parameters $\beta$ and $\mu$, independently from the initial model since the process is conducted at a later stage. In a previous analysis of LFR-CM (Orman and Labatut 2009), it has been shown that variations of the exponent value of the community size distribution $\beta$ in a realistic range have a negligible effect on the uncontrolled properties. For this reason, we used only $\beta = 2$ for the power law exponent of the community sizes distribution. On the contrary, the mixing coefficient $\mu$ is known to be the most influential parameter. We consequently used many different values, making it range from $0.05$ to $0.95$ with a $0.05$ step.

EV additionally allows controlling the transitivity, and we found out score $b = 1.5$ and selection pressure $\varepsilon = 0.99$ gave the highest values. Using the method described in section 3, we generating three benchmarks, by producing 25 networks for each one of the three models, using each combination of parameters.

*5.2 Generated Networks Properties*

In this section, we present the uncontrolled topological properties of the generated networks and discuss their realism and how they are affected by the generative model. Figure 2 shows the values obtained for the average distance, degree correlation and transitivity. Results were very similar for $\langle k \rangle = 15$ and 30, so we only present the latter here, but comments apply to both. The largest communities in the generated networks have around 700 nodes, so communities are supposed to be structurally well-defined for $\mu < 0.86$. Beyond this limit, represented on the plots under the form of a vertical line, properties values have little interest because the generated networks have no community structure, as explained in section 3.1. It is important to assess the effect of $\mu$ on the level of realism, because if we want to study the effect of the level of realism on community detection, then this level should be the more stable possible relatively to changes in the community structure.

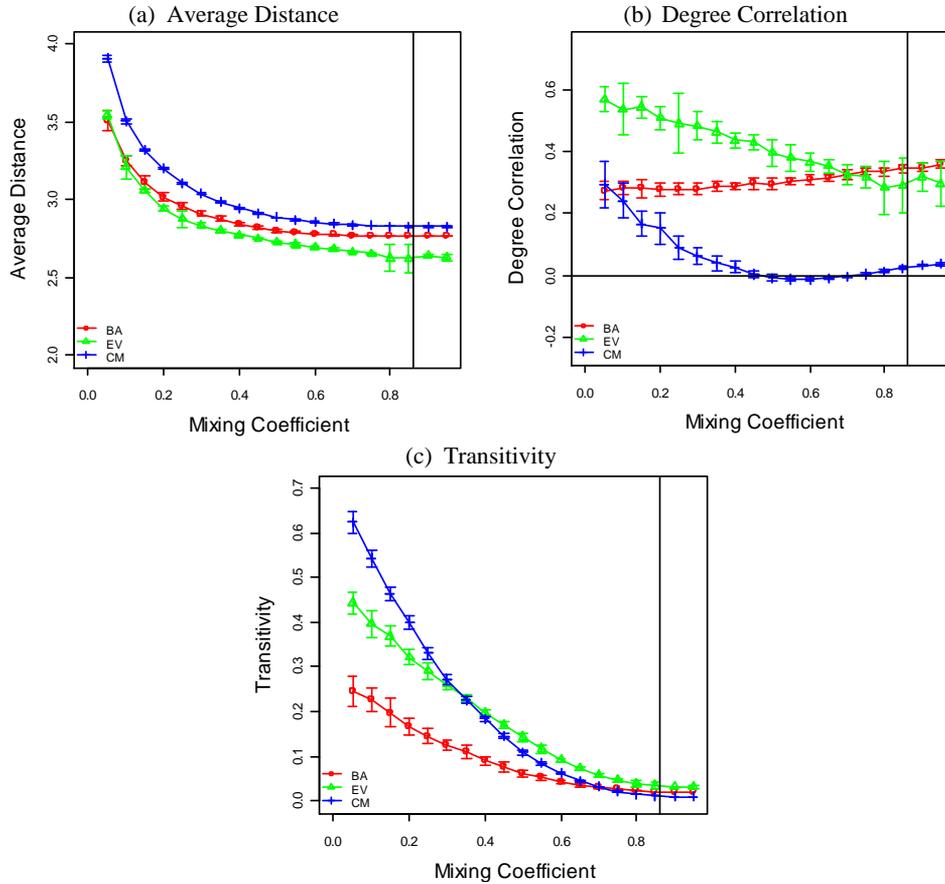

**Figure 2.** Influence of the mixing coefficient $\mu$ on the uncontrolled properties. Networks were generated with parameters $n = 5000$, $\gamma = 3$, $\beta = 2$, $\langle k \rangle = 30$ and using the three variants of the LFR method. Each point corresponds to an average over 25 generated networks. The vertical lines at $\mu = 0.86$ represent the average limit above which communities stop being clearly defined.

The average distance behaviour is rather similar for all three models. Nevertheless, its values are always slightly lower for LFR-BA and LFR-EV than for LFR-CM. It decreases monotonically as $\mu$ increases, then displays an asymptotic behaviour around $\mu = 0.3$. Indeed, for small values of the mixing coefficient, communities are well separated, i.e. there are few links between them. The paths between nodes belonging to different communities are therefore longer. As the number of links between communities increases, the shortest path lengths also decrease until the influence of community structure is negligible. The asymptotic behaviour is interesting for us, because it means algorithms compared on networks with different $\mu$ will meet the same level of small-worldness.

LFR-CM has the highest transitivity, with values around 0.6 for $\mu \approx 0$, but it also has almost zero transitivity for $\mu_{lim}$, exhibiting a serious sensitiveness to $\mu$. Both other models also show a decreasing transitivity when $\mu$ increases, but the range is much smaller, mainly because their values for $\mu \approx 0$ are significantly smaller: around 0.25 and 0.45 for LFR-BA and LFR-EV, respectively. Like LFR-CM, they reach almost zero for $\mu_{lim}$. So contrarily to what we expected, networks generated with LFR-EV do not have a higher transitivity than LFR-CM, at least for low $\mu$ values. However, thanks to its lesser sensitivity to $\mu$, LFR-EV has a better transitivity for $\mu > 0.3$. Note that in the literature, real-world networks with a transitivity greater than 0.3 are considered highly transitive (da Fontura Costa et al. 2008), so we can state all three models exhibit realistic transitivity for low $\mu$ values. The issue is more about their sensitivity to the mixing coefficient $\mu$, leading to non-realistic values for high $\mu$ values. This behaviour observed for all three models could be linked to the rewiring process performed by the LFR method.

Considering the degree correlation, there is a clear difference between LFR-CM and the other two models. LFR-CM generates networks with realistic degree correlation values for well separated communities but this value decreases rapidly and oscillates around zero for $\mu > 0.4$. LFR-EV exhibits the highest degree correlation, with values greater than 0.5 for $\mu \approx 0$. It also decreases linearly when $\mu$ increases, resulting in values close to 0.25 for $\mu \approx 1$. Finally, unlike other models, LFR-BA degree correlation increases linearly with $\mu$, ranging approximately from 0.25 ($\mu \approx 0$) to 0.35 ($\mu \approx 1$). It is also noteworthy that the statistical variations for this model are much lower than for the two others.

Figure 3 displays the evolution of the different centralisations. For the betweenness centralization, the values are very close to zero, and very stable for all models. This means the generated networks do not contain any critical node which would be lying on numerous shortest paths. For the two other centralisations, the obtained values are always higher for LFR-BA and LFR-EV than for LFR-CM, meaning the network structures are different. Moreover, for the latter the values are very close to zero, which indicates this model does not produce networks containing influential nodes. The higher values observed for both LFR-BA and LFR-EV may be linked to the preferential attachment process used in these models. It tends to generate hubs, whose presence increases the degree and closeness centralisation, as shown by our results. From the magnitude of the measured values, we can confirm the generated networks contain at least a few hubs. From the evolution of the degree centralisation, we can note the rewiring process only slightly affects how central the most central nodes are. This is due to the fact the LFR rewiring step preserves the degree distribution.

To summarize, we can state LFR-BA and LFR-EV produce more realistic networks than LFR-CM, in the sense their topological properties are closer to those encountered in real-world networks. Their average distance is lower, their degree correlation is higher, and their centralisation is higher. Networks generated with LFR-CM are nevertheless more transitive, at least when the communities are well separated. This advantage is reduced as their transitivity decrease faster than for the both other models when the mixing coefficient increases. Between LFR-BA and LFR-EV, the latter has better average distance, degree correlation and transitivity, while the former is more centralized. In terms of sensitivity to $\mu$, all models display comparable behaviours on the average distance, betweenness and degree centralisations. However, on the degree correlation, transitivity and closeness centralisation, LFR-BA is the more stable, followed by LFR-EV, while LFR-CM is the most sensitive. We can conclude both proposed variants of the LFR method offer improved realism and stability, each one presenting different advantages. The next subsection will be dedicated to study how these differences in stability and realism translate in terms of community detection performances.

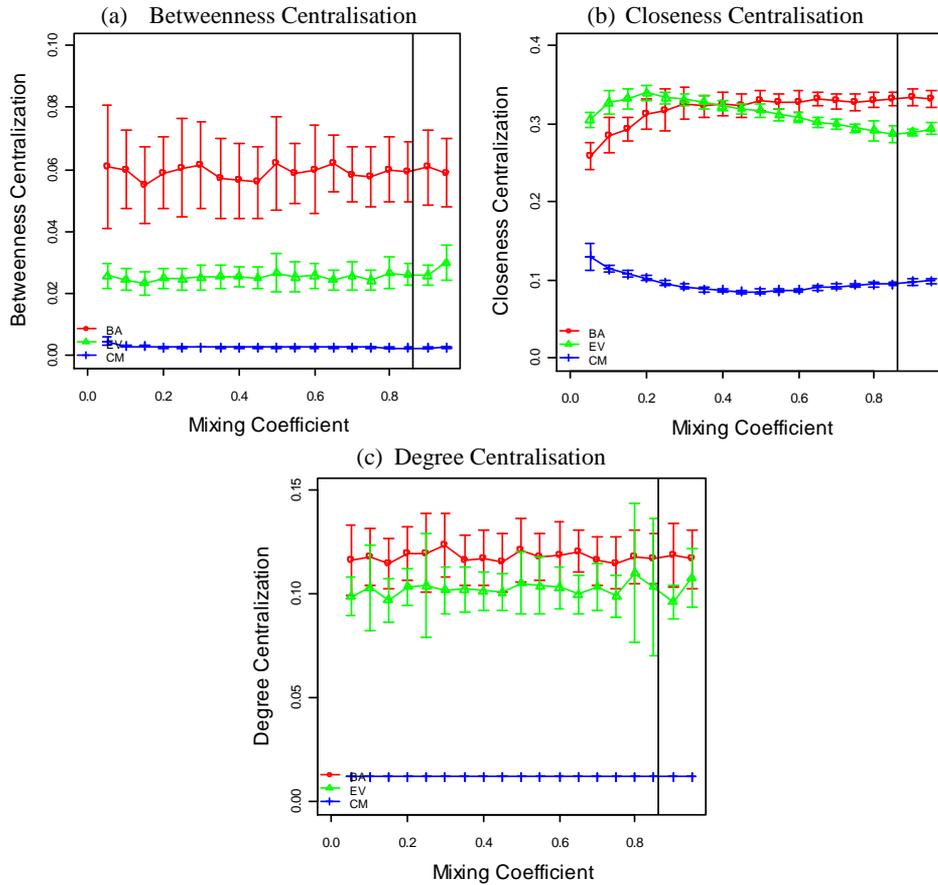

**Figure 3.** Influence of the mixing coefficient $\mu$ on the centralisation measures. Networks were generated with parameters n = 5000, $\gamma$ = 3, $\beta$ = 2, ⟨k⟩ = 30 and using the three LFR variants.. Each point corresponds to an average over 25 generated networks. The vertical lines at $\mu = 0.86$ represent the average limit above which communities stop being clearly defined.

*5.3 Community Detection Performances*

The community detection algorithms presented in section 4 have been applied to all the generated networks. To measure their performances, we used the normalized mutual information (NMI) as it is commonly used in the community detection literature (Danon et al. 2005; Lancichinetti et al. 2008; Lancichinetti and Fortunato 2009a, b). We treated the networks with average degree $\langle k \rangle = 15$ and 30, but did not observe any relevant difference between their results: the performances were just uniformly slightly better for 30 than for 15. Consequently, our plots from Figure 4 show only the former. In the rest of this section we study the effect of two factors on the algorithms performance: the level of separation of the communities, measured by the mixing coefficient $\mu$, and the level of realism of the networks, which depends directly on the generative model.

Generally, as expected from previous studies (Lancichinetti et al. 2008; Lancichinetti and Fortunato 2009a, b; Orman and Labatut 2009), the accuracy of all algorithms decreases when $\mu$ increases, i.e. as communities become more mixed and difficult to identify. Overall, we can distinguish three classes of behaviours depending on the evolution of these performances.

The algorithms from the first class all manage to perfectly identify the community structures for low mixing coefficient values, and for all three models. This class contains InfoMap, LabelPropagation, Louvain, MarkovCluster, SpinGlass and WalkTrap (the first two rows in Figure 4). For these algorithms, and when $\mu$ is small enough, we do not observe any difference between the three benchmarks. In other words, the level of realism of the networks has no influence on their performances. When the mixing coefficient increases, however, their performances deteriorate in a sharp way and some differences between the various benchmarks appear. We can order the algorithms in terms of robustness against the variations induced by the three generative models, by comparing the $\mu$ values corresponding to these points of divergence. The less sensitive is InfoMap, followed by SpinGlass and WalkTrap. For them, performances start diverging approximately when $\mu$ exceeds 0.6. Louvain starts to be sensitive to the model when $\mu$ reaches 0.5. For MarkovCluster and LabelPropagation, the differences appear when the $\mu$ is around 0.2. Note that LabelPropagation is the most sensitive to this model effect, not only because its point of divergence appears very soon, but also because its divergence is the wider. Furthermore, it is very sensitive to statistical fluctuations, as indicated by its dispersion bars. Except for LabelPropagation, the lowest performances are always obtained on the LFR-BA networks. There are very slight differences for SpinGlass, WalkTrap and Louvain between the LFR-CM and LFR-EV networks. The performances are higher with LFR-CM for InfoMap and LabelPropagation, while MarkovCluster performs better on the LFR-EV networks.

For the remaining algorithms, there is always a difference of performance due to the model, whatever the considered mixing coefficient. Nevertheless, one can distinguish two distinct classes thanks to the general shape of the performance curves. The second class includes InfoMod, Radetal, and Leading EigenVector (third row in Figure 4), in decreasing order of efficiency. Those algorithms manage to have relatively stable performances for low mixing coefficient values, although those are not optimal like for the first class. Then at some point, those performances decrease and get close to zero. Infomod and Leading Eigenvector obtain better results on LFR-CM networks, then LFR-EV and finally LFR-BA. Radetal has a very atypical behaviour, as there is no clear ordering of its performances relatively to the three models.

The third class contains CommFind and FastGreedy (last row in Figure 4). For both algorithms, the performances decrease almost linearly as soon as the mixing coefficient increases. For FastGreedy, the differences observed between the models are not statistically significant, so we can conclude it is not sensitive to the realism of the networks. For CommFind, the results are very similar for LFR-CM and LFR-EV, but the performances deteriorate much faster for LFR-BA. Note the performances are well below the other algorithms, for both CommFind and FastGreedy, especially when the mixing coefficient exceeds 0.2.

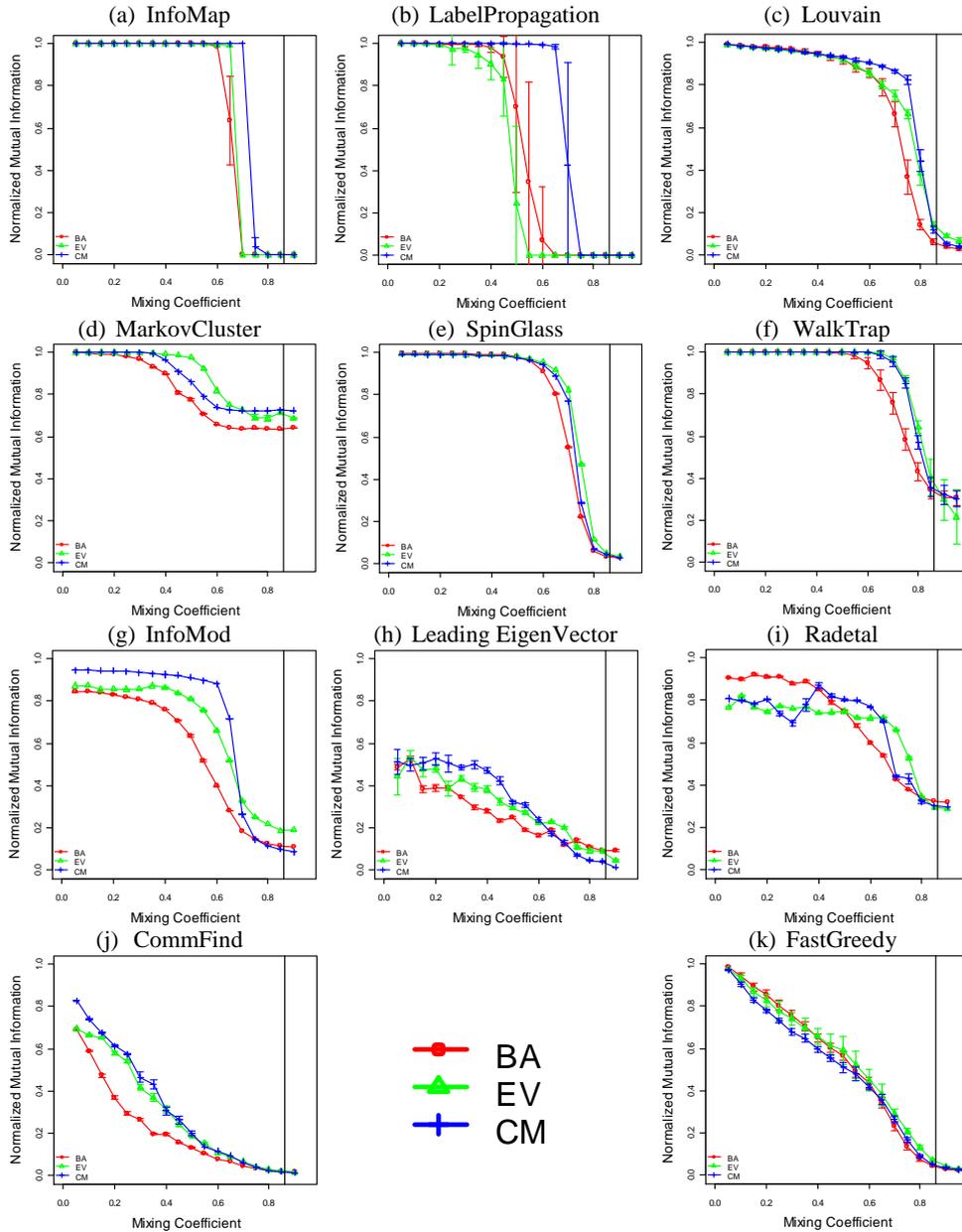

**Figure 4.** Performance of the community detection algorithms. Networks were generated with parameters $n = 5000$, $\gamma = 3$, $\beta = 2$, $\langle k \rangle = 30$ and using the three variants of the LFR method. Each point corresponds to an average over 25 generated networks. The vertical lines at $\mu = 0.86$ represent the average limit above which communities stop being clearly defined.

From these results, we can conclude the principle behind the community detection process is not a relevant factor to discriminate the algorithms relatively to how they respond to network realism. Algorithms based on very different methods can exhibit similar behaviours, whereas others relying on the same approach can obtain radically different results. For example, InfoMap and InfoMod are both information theory-based, but InfoMap displays almost no sensitiveness to the model effect, whereas InfoMod does very much. In general, as long as the proportion of inter-community links is reasonable, the most efficient algorithms are hardly influenced by the level of realism. However, when the boundaries between communities get fuzzier, this factor becomes significant. Globally, LFR-BA networks are the most difficult to process, whereas those generated by LFR-CM are associated to the highest performances. The LFR-EV model lies somewhere in between. It is important to notice that, for values of $\mu$ exceeding the points of divergence identified for the algorithms of the first class, LFR-BA displays the most realistic properties, or properties similar to the other models. This is due to its stability relatively to $\mu$, and is valid for all the studied properties. We can consequently say the lowest performances are obtained for the most realistic networks, which confirms previous studies regarding the effect of network realism on community detection.

## 6 Conclusion

In this article, we investigated the effects of the realism level of artificially generated complex networks on the performance of community detection algorithms. We based our work on the LFR generative method, which relies itself on the configuration model (CM). In order to improve the realism of the networks it produces, we proposed to substitute the Barabási–Albert (BA) and the evolutionary preferential attachment (EV) models to the CM. We generated three distinct benchmarks (LFR-CM, LFR-BA and LFR-EV) and studied their topological properties. It turns out both proposed modifications lead to more realistic networks in terms of average distance, degree correlation and centralisation. For the first three properties, LFR-EV globally exhibits better absolute values. But LFR-BA has higher centralisations, and is less sensitive to the level of separation of the communities.

We applied a wide range of community detection algorithms on each benchmark, in order to analyse the effect of these modifications. Overall, the performances decrease when the realism of the networks increases. In general, the best results are obtained for LFR-CM, whereas LFR-BA leads to the lowest performances. We distinguished three classes of algorithms depending on their results. In the first, differences between models appear only when the proportion of intercommunity links is high enough to make the community detection problem a difficult task. Among these algorithms, InfoMap displays the highest performances, followed by SpinGlass and WalkTrap. The algorithms from the second and the third class are always sensitive to the benchmarks variations whatever the proportion of the inter community links. The shape of their performance curves is the main characteristic allowing to distinguish them. Those of the second class show stable performances, although not as good as the first class, for clearly separated networks; then their performances drop quickly. The performances of the algorithms of the third class decrease monotonically when the proportion of inter-community links increases.

Amongst the three models, LFR-BA is the most appropriate for the evaluation of community detection algorithms. Compared to LFR-CM, the topological properties of the networks it generates are more stable relatively to changes in the community structure. This is also the case for LFR-EV, but LFR-BA is more stable. This stability allows performing consistent comparisons: networks can have more or less separated communities, while attaining approximately the same level of realism. This level is also higher for both LFR-BA and LFR-EV, compared to LFR-CM, since the obtained topological properties are closer to what is encountered in real-world networks. LFR-EV obtains even more realistic values for average distance, degree correlation and transitivity, but LFR-BA makes up for this thanks to its better stability. Indeed, LFR-BA gets very close to LFR-EV for these properties when the proportion of links between communities reaches a certain level. In other words, when community detection becomes difficult, LFR-BA is either as much, or more realistic than LFR-EV. This means the networks it produces are particularly adapted to the evaluation of the first class algorithms.

The modifications we proposed were efficient to improve the realism of the networks generated by the LFR method. Nevertheless, they also resulted in a loss of control, since the replacement models (BA and EV) do not allow to specify directly as many properties as CM, such as the exponent of the degree power law distribution. Different ways can be explored to try to solve these limitations. First, it would be interesting to study the side effects of the rewiring process used in the LFR approach, by simply comparing the generated networks properties before and after the rewiring step. This work is necessary to determine if some properties observed in the final networks depend on the initial (pre-rewiring) network or on the rewiring process itself. Second, many other models exist to generate networks with a power law distributed degree (Chen and Chen 2007; Chuang et al. 2009; Tam et al. 2008). A systematic review could allow detecting more flexible models, offering more control on the generated networks properties, and more realistic properties.

**Acknowledgement**

Part of this study has been presented at the International Conference on Advances in Social Networks Analysis and Mining (ASONAM 2010) as a short paper (Orman and Labatut 2010). This work was partially funded by the AUF (Association of French-speaking Universities).